\documentclass[12pt]{article}
\usepackage[dvips]{graphicx}

\newcommand{\GeV}{\ensuremath{\mathrm{Ge\kern -0.12em V}}}
\newcommand{\TeV}{\ensuremath{\mathrm{Te\kern -0.12em V}}}

\def\wangms{\mathrm{sin^{2}\theta_{\overline{MS}}}}
\def\cwangms{\mathrm{cos^{2}\theta_{\overline{MS}}}}
\def\wangmsmz{\mathrm{sin^{2}\theta_{\overline{MS}}(M_Z)}}

\def\qq{\mathrm{q\bar{q}}}

\begin{document}
\begin{center}

{\Large\bf Strong Coupling Running, Gauge Coupling Unification and the Scale of New Physics}
\vspace{1cm}

{\sc 
Dimitri Bourilkov
}

{\small bourilkov@mailaps.org
}

{\sl
Physics Department, University of Florida, P.O. Box 118440\\
Gainesville, FL 32611, USA}

\end{center}

\begin{abstract}
The apparent unification of gauge couplings in Grand Unified Theories
around 10$^{16}$~$\GeV$ is one of the strong arguments in favor of
Supersymmetric extensions of the Standard Model. In this paper, an
analysis of the measurements of the strong coupling running from the
CMS experiment at the LHC is combined with a ``traditional'' gauge
coupling unification analysis using data at the Z peak. This approach
places powerful constraints on the possible scales of new physics and
on the parameters around the unification scale. A supersymmetric
analysis without GUT threshold corrections describes the CMS data well
and provides perfect unification. The favored scales are
\mbox{$M_{SUSY}\ =\ 2820\ +670\ -540\ \GeV$} and
\mbox{$M_{GUT}\ =\ 1.05 \pm 0.06 \cdot 10^{16}\ \GeV$}.
For zero or small threshold corrections the scale of new physics may
be well within LHC reach.
\end{abstract}

\section{Introduction}

The Standard Model (SM) of particle physics works extremely well at
collider energies. But it is incomplete, and the possible scale at which
new physics phenomena could manifest themselves is a hot topic of
great theoretical, experimental and practical interest.

When the LEP and SLC colliders came online, the weak coupling was measured
with much higher precision, roughly on par with the precision for the
electromagnetic coupling. In a renowned paper~\cite{Amaldi} from 1991
the famous plot was produced, showing that in contrast to the SM the
Minimal Supersymmetric Standard Model (MSSM) leads to a single
unification scale of a Grand Unified Theory (GUT), if we let the
couplings run according to the MSSM theory:
\begin{equation}
          M_{SUSY} = 10^{3.0 \pm 1.0} \GeV \ ;\ 
          M_{GUT}  = 10^{16.0 \pm 0.3} \GeV \ ;\ 
          1/\alpha_{GUT} = 25.7 \pm 1.7
\end{equation}
where $M_{SUSY}$ is a single generic SUSY scale where the spectrum
of supersymmetric (SUSY) particles starts to play a role by changing the
running of the couplings, $M_{GUT}$ is the scale of grand unification
where the electromagnetic, weak and strong coupling come together as
an unified coupling $\alpha_{GUT}$.

In this paper indirect analyses based on the running of gauge couplings
are performed to search for the scale of Supersymmetry, or any new
physics that could change the running of the couplings. Two analyses
are carried out:
\begin{itemize}
 \item first a - by now ``traditional''~\footnote{See e.g.~\cite{Amaldi}.}
  - analysis of gauge coupling unification in a grand unification theory,
  using the latest experimental inputs at the Z peak scale, combined with
  a detailed statistical approach
 \item a novel analysis, including for the first time the measurements of
  the strong coupling by the CMS collaboration using the data collected
  at 7 $\TeV$~\cite{Chatrchyan:2013txa,Khachatryan:2014waa,CMS:2014mna},
  which can shed light on the actual running of couplings from the Z peak
  to $\approx$~1.4 $\TeV$.
\end{itemize}

\section{Digression}

If we are looking at the front of a train, we can see the lead
locomotive but cannot tell much about the composition, or even how
many locomotives are hauling it. If we are on the railway platform, we
have a much better view.

In the first approach outlined above, in contrast to the SM the
additional degree of freedom provided by the introduction of a new
MSSM scale could lead to a single GUT unification scale, if we let the
couplings run accordingly. It is amazing that we are trying to
extrapolate from 10$^2$ to 10$^{16}$ $\GeV$. From experiments we know
that even interpolation or modest extrapolations can be non-trivial.
We measure the ``offsets'', i.e. the values of the three couplings,
around the Z peak and rely on theory to give us the ``slopes'' of the
running couplings - without errors - up to the GUT scale. This may be
an illusion e.g. new physics phenomena like extra dimensions could
modify the running at intermediate scales, so the ``unification''
point may be imaginary.

In the second approach, we step on the railway platform. The actual
running of the strong coupling provides a powerful constraint on new
physics scenarios.

\section{Definitions and Experimental Inputs}

Throughout this paper the couplings are denoted as $\alpha_i$, $i\ =\ 1,2,3$.
The values of $1/\alpha$ are used for plots and fit results.
For the renormalization the modified minimal subtraction scheme
($\overline{MS}$) is used. The definitions are:
\begin{eqnarray}
\alpha_1 = \frac{5}{3}\frac{\alpha}{\cwangms}\\
\alpha_2 = \frac{\alpha}{\wangms}\\
\alpha_3 = \frac{g_s^2}{4\pi}
\end{eqnarray}
where $\alpha$ is the electromagnetic coupling, $\wangms$ is the
electroweak mixing angle in the $\overline{MS}$ scheme, and
$\alpha_3$ is the strong coupling.

From Review of Particle Properties 2014~\cite{RPP2014}
(sections 10.2, 10.6, 9.3.12):
\begin{eqnarray}
1/\alpha(M_Z) = 127.940 \pm 0.014\\
\wangmsmz = 0.23126 \pm 0.00005\\
\alpha_{s}(M_Z) = 0.1185 \pm 0.0006.
\end{eqnarray}

In 1991, at the Z mass scale $M_Z$, the relative errors in
$\alpha(M_Z)$, $\wangmsmz$ and $\alpha_{s}(M_Z)$ were 0.24, 0.77 and
4.6 \% respectively.

Now they are 0.011, 0.022 and 0.5 \%, so we have improved enormously
in the last quarter of a century. Even the measurements of the
notoriously hard to extract strong coupling have gained an order of
magnitude in precision. All this helps to improve the precision of the
unification analyses.

In addition, for the second analysis, three measurements of the strong
coupling by the CMS collaboration using the data collected at 7 $\TeV$
are utilized. They are using the ratio of the inclusive 3-jet cross
section to the inclusive 2-jet cross section
$R_{32}$~\cite{Chatrchyan:2013txa}, inclusive jet cross sections up to
2 $\TeV$ in jet transverse momentum~\cite{Khachatryan:2014waa}, and
inclusive 3-jet production differential cross sections, where the
invariant mass of the three jets is in the range of 445--3270
$\GeV$~\cite{CMS:2014mna}. The Q values at which the strong coupling
is measured are (474, 664, 896)~$\GeV$, (136, 226, 345, 521, 711, 1007)~$\GeV$
and (361, 429, 504, 602, 785, 1164, 1402)~$\GeV$ respectively. The 3-jet
measurement reaches the highest scales.

\section{Analysis Technique}

A $\chi^2$ minimization with {\tt MINUIT} for three parameters:
$M_{SUSY}$, $M_{GUT}$ and $1/\alpha_{GUT}$ is used for the
``traditional'' analysis. The three couplings at $M_Z$ serve as input.
The procedure is similar to the one used
in~\cite{Bourilkov:2004zu,Bourilkov:2006rz}.

In the case of 1-loop Renormalization Group (RG) running of the
couplings we can solve analytically; the coefficients for the 3
couplings are independent, given by the following SM or MSSM equation
relating the couplings at scales $\mu$ and $\nu$:
\begin{equation}
1/\alpha_i(\nu) = 1/\alpha_i(\mu) - (b_i/2\pi) \cdot ln(\nu/\mu) .
\end{equation}

In the SM above the top threshold the $b_i$ coefficients are
\mbox{(4.1, -3$\frac{1}{6}$, -7)}, while in the MSSM they have the
values \mbox{(6.6, 1, -3)}. As can be seen, the changes in slopes are
substantial for all three couplings.

The high precision of the experimental inputs requires to use
2-loop RG running. In this case additional non-diagonal terms
$b_{ij}$~\mbox{(i,j = 1,2,3)}
appear in the equations, so the 3 couplings depend on each other and
the error propagation is altered. The equations are solved
numerically. Now the errors depend on the scale - for typical GUT
scales they grow by 4, 12 and 6 \% respectively for the three
couplings.

The details of physics at the GUT scale are unknown. It may be
necessary to apply a threshold correction to $\alpha_{3}(M_{GUT})$ in
order to meet GUT boundary conditions. For detailed discussions of the
nature, sign and size of these corrections we refer the reader to the
reviews~\cite{Polonsky:2001pn} and~\cite{RPP2014}~(section~16.1.4) and
references therein.

From a technical point of view, the threshold correction is applied
as follows:
\begin{eqnarray}
\alpha_1(M_{GUT}) = \alpha_2(M_{GUT}) = \alpha_{GUT}\\
\alpha_3(M_{GUT}) = \alpha_{GUT}\cdot(1\ +\ \varepsilon_{GUT})
\end{eqnarray}
where $\varepsilon_{GUT}$ is the threshold correction for the
strong coupling at the GUT scale. In practice, the higher precision
of $1/\alpha_1$ and $1/\alpha_2$ brings the GUT scale very close
to their cross-point. The $1/\alpha_3$ line may ``hit'' or ``miss''
the unification point, depending on the situation.

In the second analysis, the $\chi^2$ minimization with {\tt MINUIT}
uses the same three parameters and adds the 16 CMS measurements,
bringing the experimental inputs to 19. In this case the fit has 16
degrees of freedom (dof).

\section{``Traditional'' Running Couplings Analysis}

I update the analyses~\cite{Bourilkov:2004zu}~\footnote{
From GUT unification a favored range of $\alpha_{s} = 0.118-0.119$,
in excellent agreement with the current world average, was derived in
this analysis.} 
and \cite{Bourilkov:2006rz}, using the same fitting procedure applied
on the new more precise experimental inputs.

The results of the fits are summarized in Table~\ref{tab:t1}.

\begin{table}[thb]
\renewcommand{\arraystretch}{1.40}
\caption{Fit results - all for 2-loop RG running.}
\vspace{0.3cm}
\begin{center}
{\begin{tabular}{@{}|c|lll|@{}} \hline
Threshold correction  & $M_{SUSY}$          & $M_{GUT}$             & $1/\alpha_{GUT}$ \\ \relax
    [\%]              & [$\GeV$]            & [$\GeV$]              &    \\ \hline
  $\pm$ 0             & $10^{3.45 \pm 0.09}$& $10^{16.02 \pm 0.03}$ & $25.83\pm 0.16$ \\
     - 3              & $10^{2.03 \pm 0.14}$& $10^{16.48 \pm 0.04}$ & $23.28\pm 0.26$ \\
     - 4              & $10^{1.48 \pm 0.11}$& $10^{16.66 \pm 0.03}$ & $22.28\pm 0.20$ \\
\hline
\end{tabular}}
\end{center}
\label{tab:t1}
\end{table}

The MSSM can still provide coupling unification at a GUT scale well
below the Planck scale for the full set of precise 2014 measurements.
But it is not without problems, which may be overlooked. As stated
in~\cite{RPP2014}~(section~16.1.4): ``A small threshold correction at
$M_{GUT}$ \mbox{($\varepsilon_{GUT} \sim -3\%\ \mathrm{to}\ -4\%$)}
is sufficient to fit the low energy data precisely.''~\footnote{
It should be noted that the experimental inputs used for the GUT
review in the Review of Particle Physics (RPP) 2014 are not the world
averages from the same review. In contrast, I have used the averages
from RPP 2014, which may explain small differences in coupling
unification fit results, but the conclusions are basically the same.}
Actually threshold corrections of this size tend to push the SUSY
scale to uncomfortably low regions, being progressively ruled out by
the LHC experiments. For a threshold correction of -4\% the fitted
SUSY scale is $M_{SUSY}$~\mbox{ = 30 +9 -7 $\GeV$}, and for a
threshold correction of -3\% the SUSY scale goes up only to
$M_{SUSY}$~\mbox{ = 107 +41 -30 $\GeV$}. In contrast a threshold
correction of 0~\% brings the SUSY scale to $\sim$~3 $\TeV$.

\section{Combined Analysis of CMS Data and Gauge Coupling Unification}

In~\cite{Bourilkov:2006rz} a fit to the precise LEP2 $\qq$ cross
section measurements above the Z pole~\cite{LEPEWWG}, with all three
couplings running simultaneously, was pioneered. In addition, the
effects of a low SUSY scale were compared to the then available
measurements of the strong coupling running above the Z - from a LEP2
combination~\cite{lepqcdwg}, and from the CDF
experiment~\cite{Affolder:2001hn}. The analysis was limited by the top
LEP2 energy or the relatively large uncertainties of the CDF result.

\begin{figure}[ht]
\centerline{\resizebox{0.95\textwidth}{10.5cm}{\includegraphics{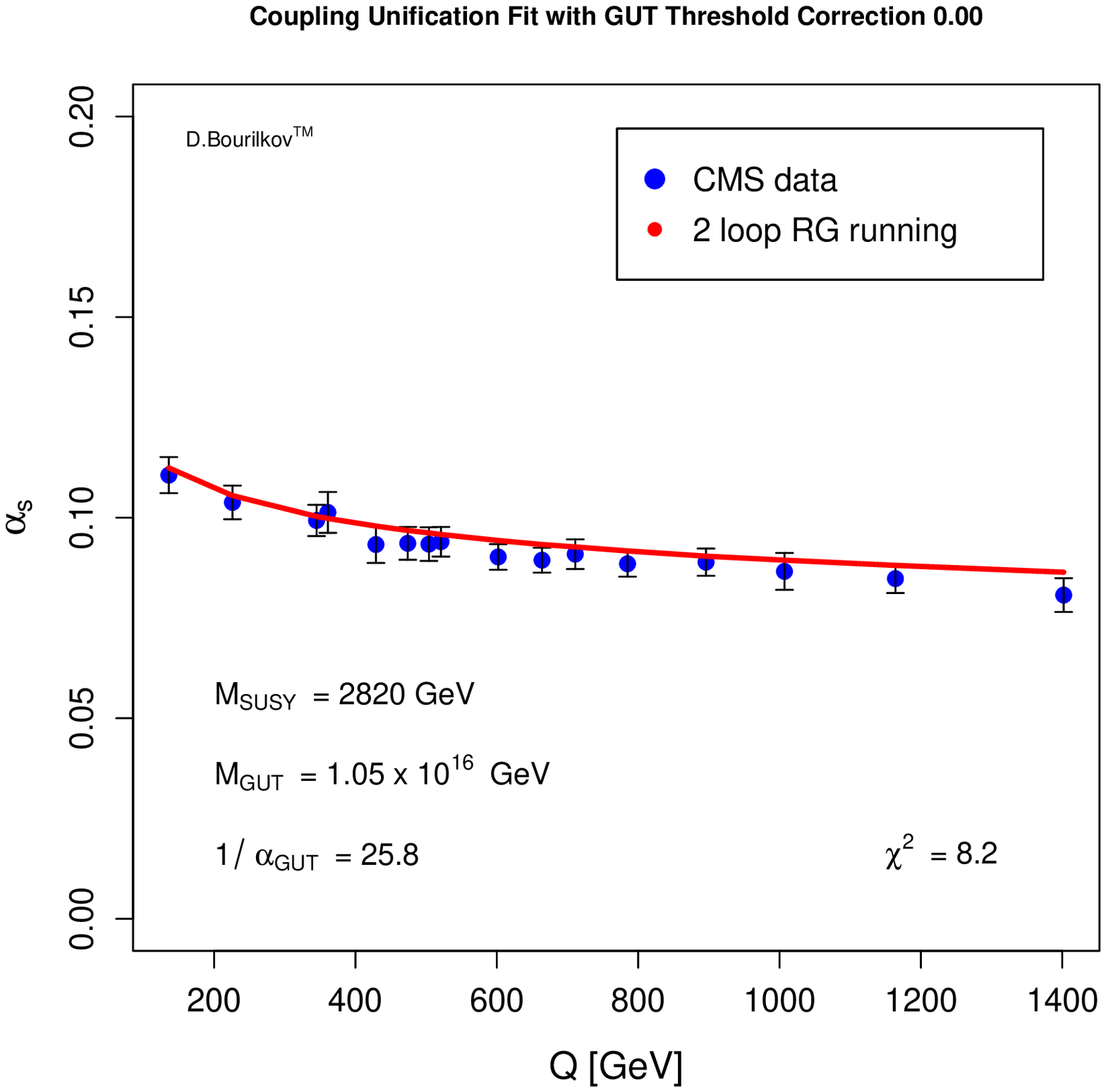}}}
\centerline{\resizebox{0.95\textwidth}{9.0cm}{\includegraphics{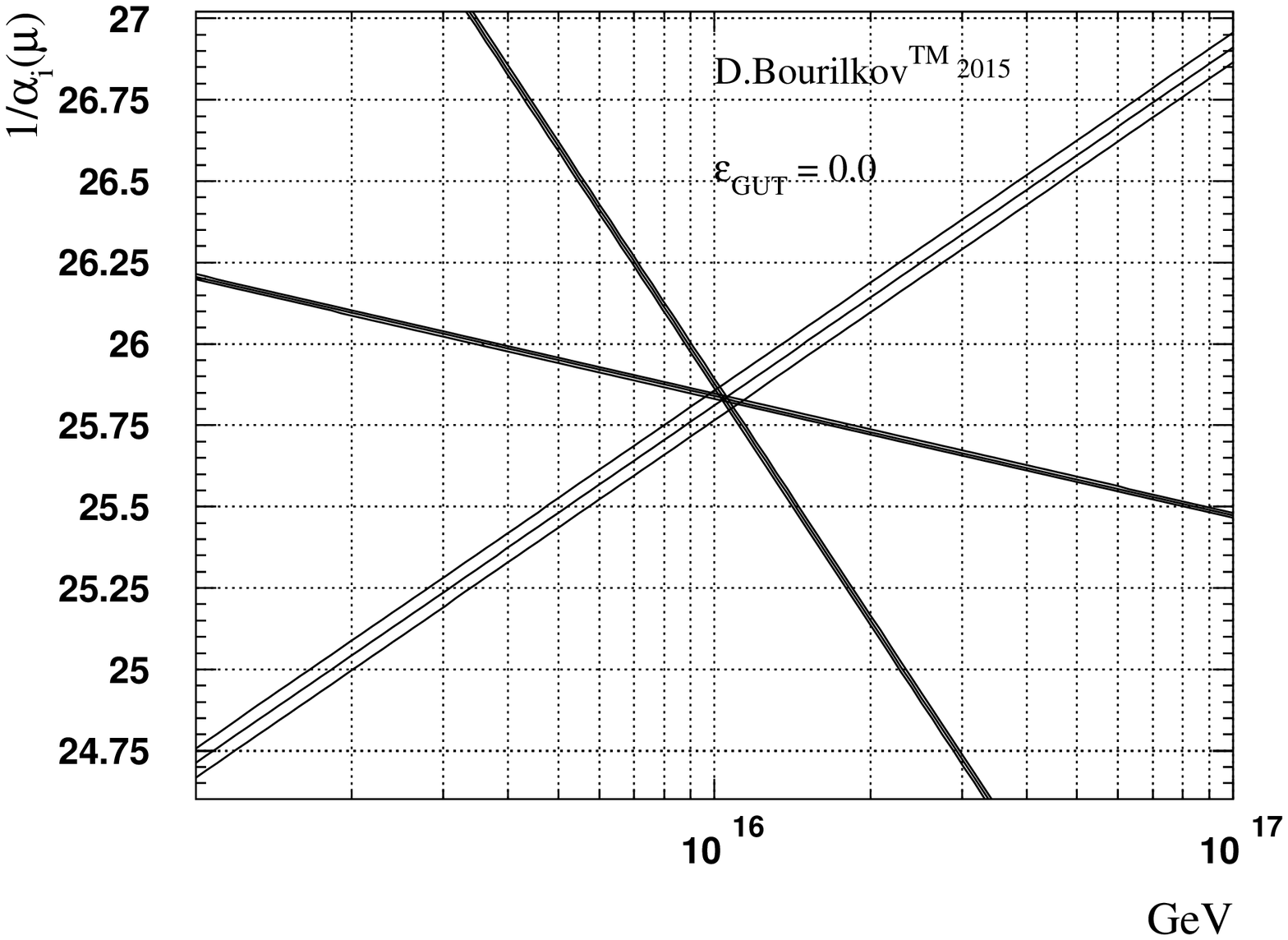}}}
\vspace*{-3pt}
\caption{Top: Coupling unification fit of the CMS $\alpha_s$ measurements with GUT
threshold correction of 0\%.
Bottom: The fitted running of the three couplings around the GUT scale.
The fit describes well the CMS measurements and works well at the GUT scale.}
\label{fig:fit000}
\end{figure}

Precision measurements of the strong coupling $\alpha_s$ well above
the Z pole can provide a new window. This coupling changes the fastest
in this energy range, but is also the hardest to measure. Recalling
the metaphor from section 2, while the above mentioned
analysis~\cite{Bourilkov:2006rz} just stepped at the end of the
railway platform, the new CMS measurements of the running of the
strong coupling in a vast area above the Z peak up to 1.4 $\TeV$ allow
for a much more central view.

A fit to the data computing the running of the couplings from the Z
peak to the highest LHC energies in the SM or in MSSM is performed.
The results for different GUT threshold corrections are summarized in
Table~\ref{tab:t2}.

\begin{table}[thb]
\renewcommand{\arraystretch}{1.40}
\caption{Combined fit results - CMS data and GUT unification.}
\vspace{0.3cm}
\begin{center}
{\begin{tabular}{@{}|c|lll|r|@{}} \hline
Threshold correction  & $M_{SUSY}$        & $M_{GUT}$           & $1/\alpha_{GUT}$ & $\chi^2$ \\ \relax
    [\%]              & [$\GeV$]         & [$\GeV$]           &                 & \\ \hline
     + 1              & $10^{3.96 \pm 0.10}$& $10^{15.85 \pm 0.03}$ & $26.74\pm 0.17$ & 8.2 \\
  $\pm$ 0             & $10^{3.45 \pm 0.09}$& $10^{16.02 \pm 0.03}$ & $25.83\pm 0.16$ & 8.2 \\
     - 1              & $10^{3.02 \pm 0.08}$& $10^{16.16 \pm 0.03}$ & $25.07\pm 0.15$ & 9.5 \\
     - 2              & $10^{2.78 \pm 0.07}$& $10^{16.25 \pm 0.02}$ & $24.63\pm 0.13$ & 25.1 \\
     - 3              & $10^{2.60 \pm 0.06}$& $10^{16.31 \pm 0.02}$ & $24.28\pm 0.10$ & 68.1 \\
     - 4              & $10^{2.42 \pm 0.05}$& $10^{16.38 \pm 0.02}$ & $23.95\pm 0.09$ & 138.3 \\
     - 5              & $10^{2.26 \pm 0.05}$& $10^{16.44 \pm 0.02}$ & $23.66\pm 0.09$ & 235.7 \\
\hline
\end{tabular}}
\end{center}
\label{tab:t2}
\end{table}

For a threshold correction of 0\%, the 2-loop RG running provides an
excellent fit for the CMS data and perfect unification, as shown in
Figure~\ref{fig:fit000}. The favored SUSY scale is 2820~$\GeV$,
which means that the $\alpha_s$ measurements are well described
by 2-loop SM running, as implemented in this analysis.

The $\chi^2/dof$ of the fit is $\sim 0.5$. This is not surprising, as
the statistical and systematic errors are combined in
quadrature.  The correlations between the systematic errors are
not provided and are not taken into account. They can be important
both within each of the three CMS measurements, and on top of this
hard to determine correlations between them could exist.

\begin{figure}[ht]
\centerline{\resizebox{0.95\textwidth}{10.5cm}{\includegraphics{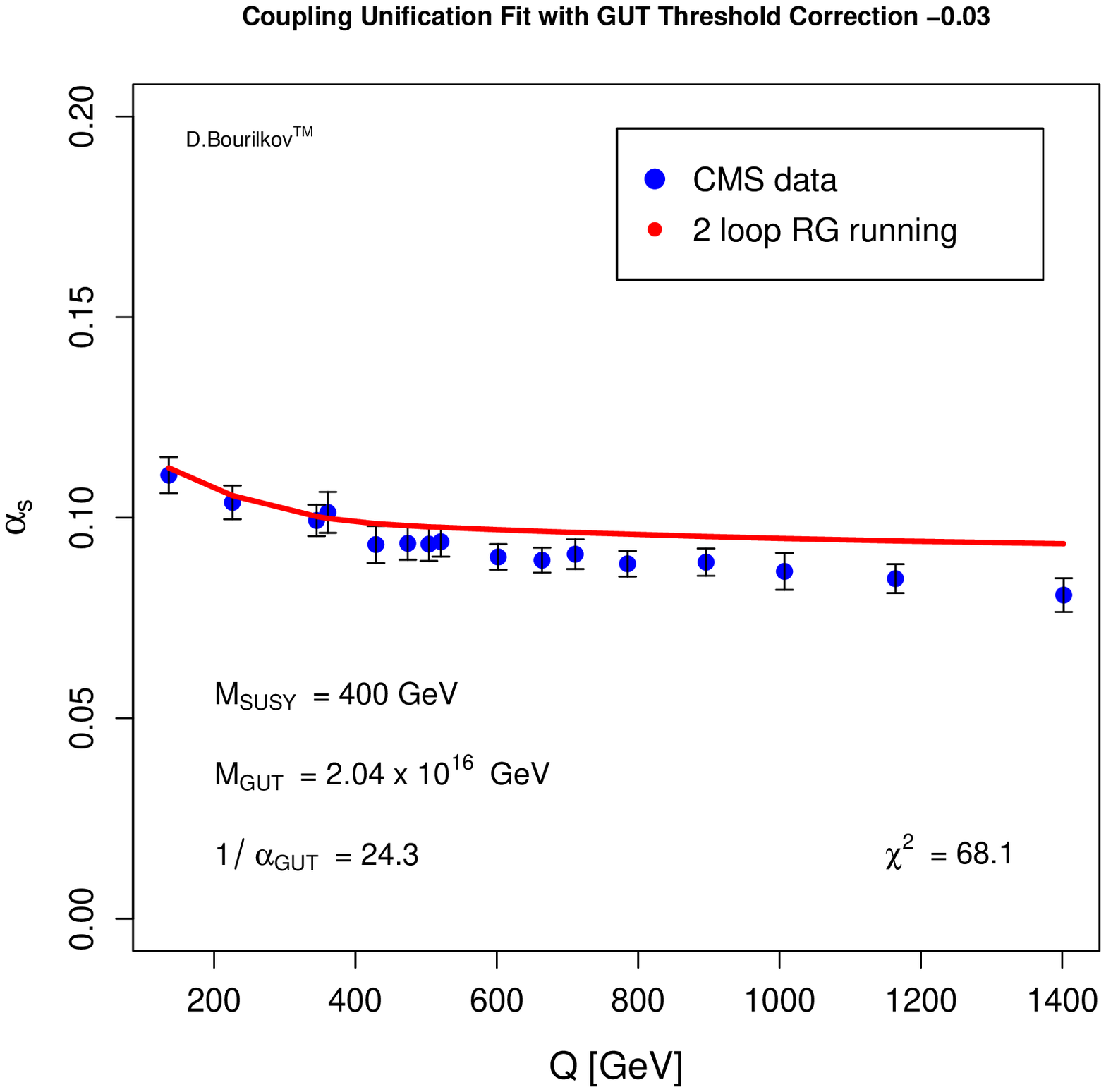}}}
\centerline{\resizebox{0.95\textwidth}{9.0cm}{\includegraphics{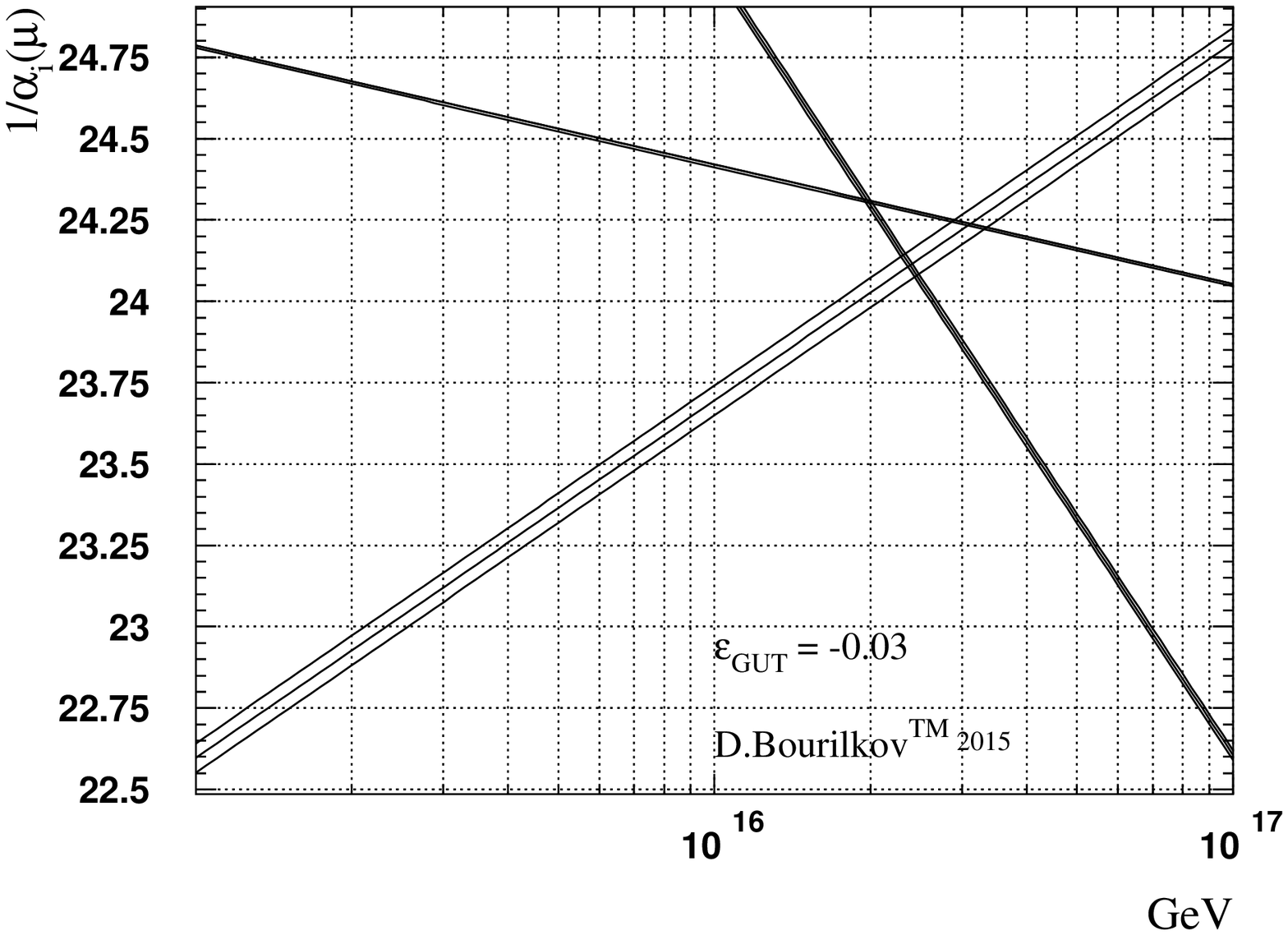}}}
\vspace*{-3pt}
\caption{Top: Coupling unification fit of the CMS $\alpha_s$ measurements with GUT
threshold correction of -3\%.
Bottom: The fitted running of the three couplings around the GUT scale.
The fit fails on both counts as evidenced by the $\chi^2$: the CMS measurements
are not well described and it does not work well at the GUT scale.}
\label{fig:fit003}
\end{figure}

\begin{figure}[ht]
\centerline{\resizebox{0.95\textwidth}{10.5cm}{\includegraphics{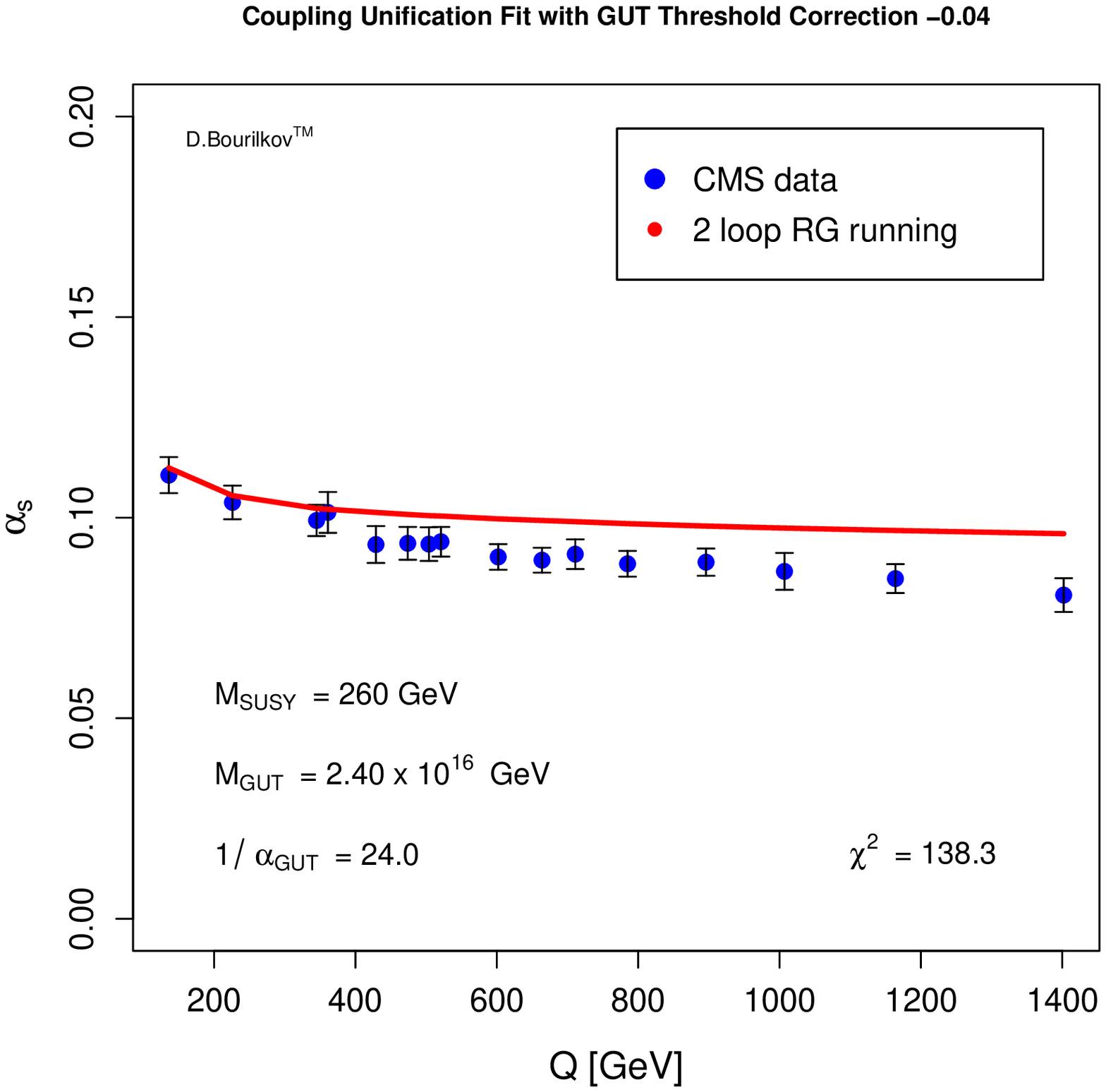}}}
\centerline{\resizebox{0.95\textwidth}{9.0cm}{\includegraphics{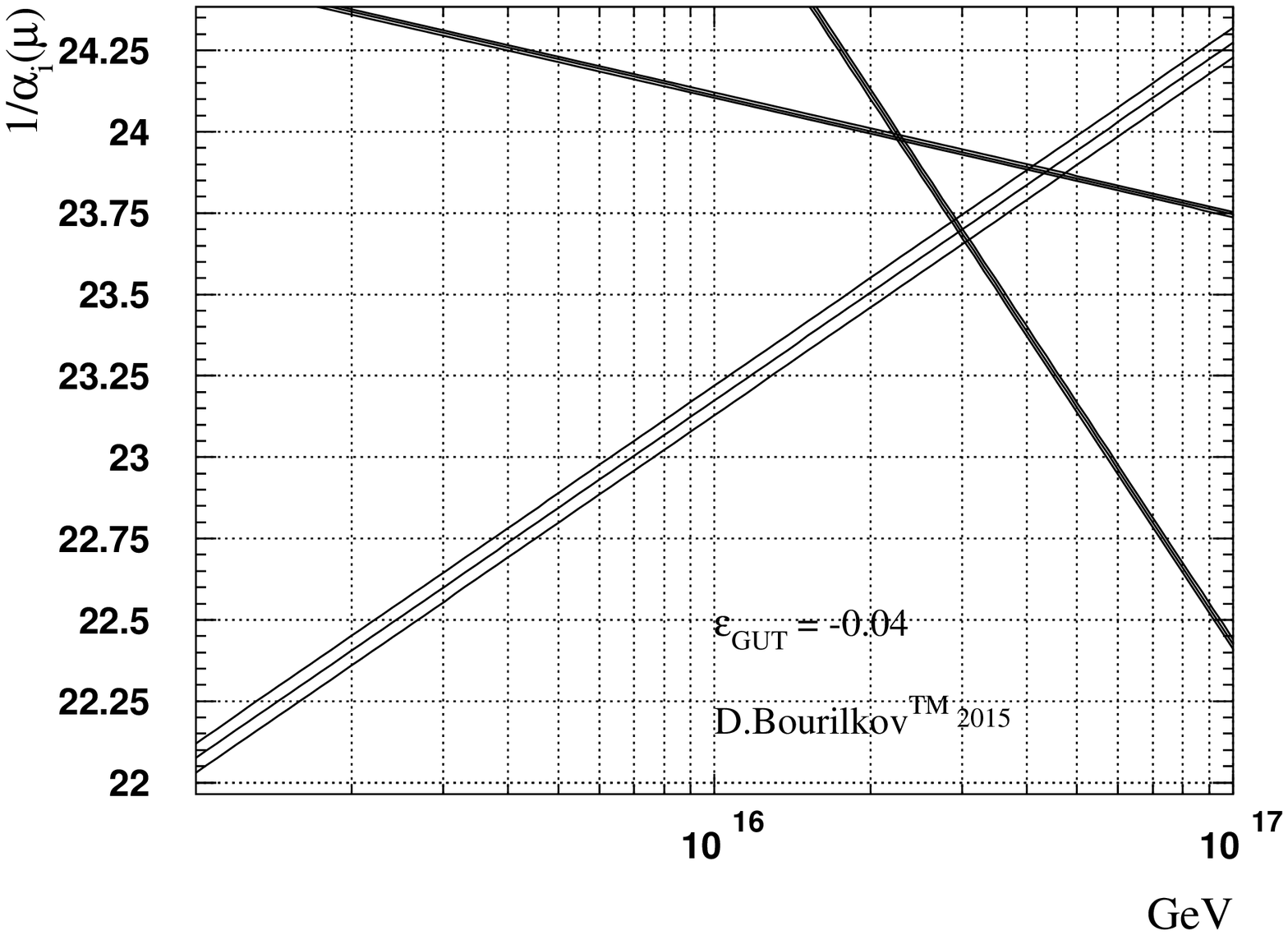}}}
\vspace*{-3pt}
\caption{Top: Coupling unification fit of the CMS $\alpha_s$ measurements with GUT
threshold correction of -4\%.
Bottom: The fitted running of the three couplings around the GUT scale.
The fit fails on both counts as evidenced by the $\chi^2$: the CMS measurements
are not well described and it does not work well at the GUT scale.}
\label{fig:fit004}
\end{figure}

For a threshold correction of -3\%, the fit runs into trouble on both
ends. The unification point tries to push the SUSY scale down to low
values (as in the ``traditional'' analysis), but this is not
compatible with the CMS data. As a result the fit ends up with an
unsatisfying compromise: a poor description of the CMS data and a
``miss'' at the GUT scale, as shown in Figure~\ref{fig:fit003}.  The
$\chi^2$ grows from 8.2 to 68.1. The situation is even more
unsatisfactory for a threshold correction of -4\%. Here the $\chi^2$
goes up to 138.7, see Figure~\ref{fig:fit004}.

A threshold correction of -1\% is still fine. Here the $\chi^2$ goes
up only to 9.5. The preferred SUSY scale in this case - 1050~$\GeV$ -
is high enough to be only weakly constrained by the CMS data. For a
threshold correction of -2\% the $\chi^2$ already reaches a value of
25.1.

In the opposite direction, a threshold correction of +1\% (similar for
higher corrections) is not impacted by the CMS data, as the SUSY scale
is pushed up to 9120 $\GeV$ while the GUT scale is pushed down to
$0.7 \cdot 10^{16}$. Positive threshold corrections may encounter different
constraints. For example, baryon number is violated in GUT theories
and in simple scenarios the proton lifetime is proportional to:
\begin{equation}
\tau_p \sim M_{GUT}^4 \cdot (\frac{1}{\alpha_{GUT}})^2 \cdot (\frac{1}{m_p})^5
\end{equation}
The favored values for the GUT scale in this case will tend to push the
proton lifetime lower, somewhat closer to the experimental limits
$\sim$~10$^{34}$ years~\cite{RPP2014}~(section~16.1.5).

\begin{figure}[htb]
\centerline{\resizebox{0.95\textwidth}{14.0cm}{\includegraphics{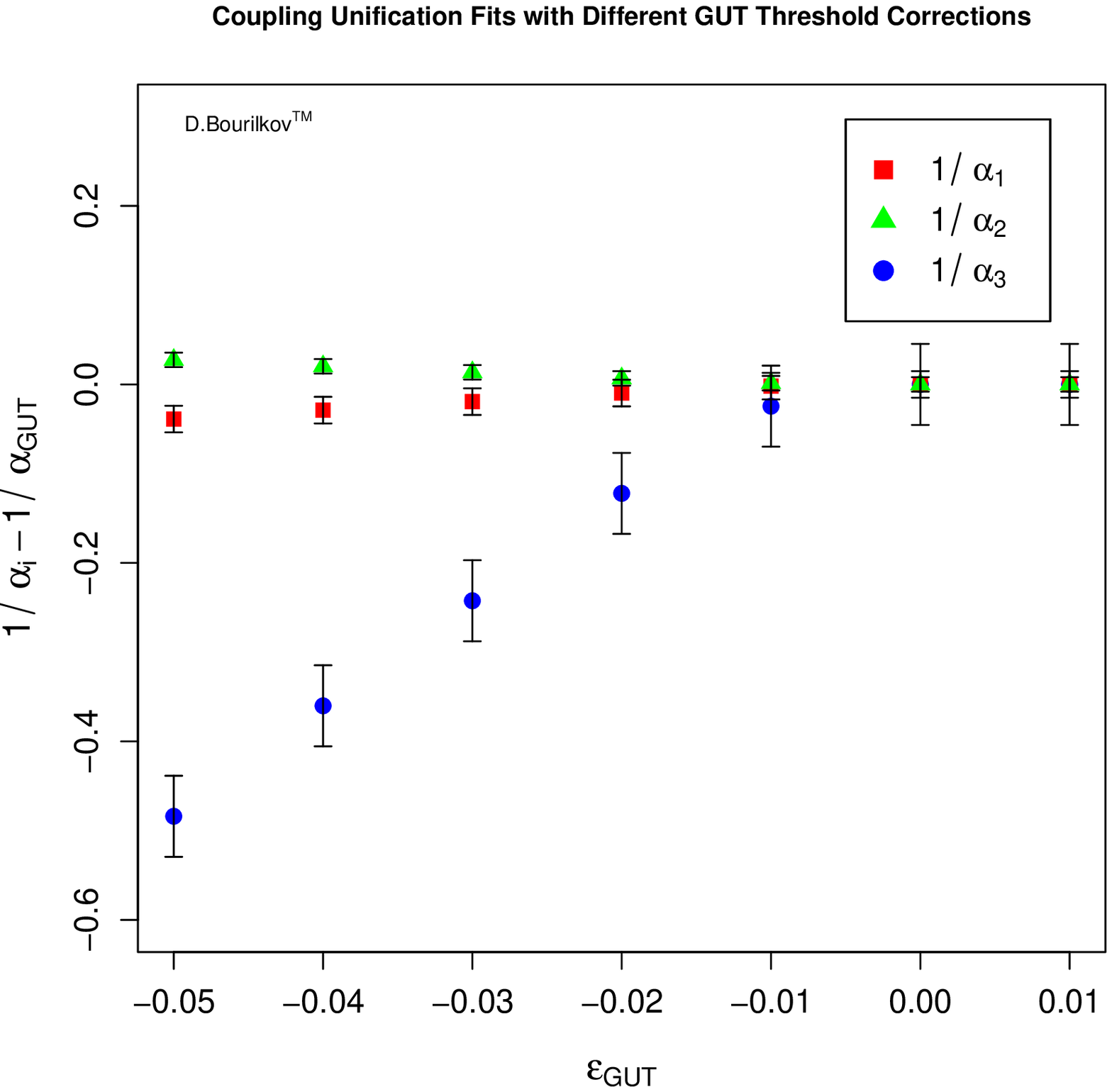}}}
\vspace*{-3pt}
\caption{Quality of the coupling unification fits for different GUT
threshold corrections $\varepsilon_{GUT}$. Clearly no or very small
threshold corrections (within $\pm 1$~\%) work best, while corrections of
$\sim$~--3-4\% fail to provide a single unification point.}
\label{fig:epsgut}
\end{figure}

The quality of the GUT unification for different threshold corrections
is visualized in Figure~\ref{fig:epsgut}.  Very small threshold
corrections, consistent with zero (or within $\pm 1\%$) work very well
and provide ``perfect'' unification.  Negative corrections quickly
diverge from a single unification point.

It is interesting to note that the results without threshold
corrections are very close to the values from the classical
analysis~\cite{Amaldi}, while the errors have improved by an
order of magnitude. The best fit for the SUSY scale gives:
\begin{equation}
M_{SUSY}\ =\ 2820\ +670\ -540\ \GeV.
\end{equation}
The corresponding range from~\cite{Amaldi} was 100--10000~$\GeV$.
For a threshold correction of -1\% the best fit is
\mbox{$M_{SUSY}\ =\ 1050\ +220\ -180\ \GeV$}.

In this paper a MSSM fit to the running couplings is used as a
baseline. Similar analyses using the CMS measurements of the running
of $\alpha_s$ can be performed for any new physics scenario affecting
the couplings, if the relevant scales are in the measurement
range. Many variations are possible. In~\cite{Becciolini:2014lya} the
CMS ratio of the inclusive 3-jet cross section to the inclusive 2-jet
cross section (with highest point 896~$\GeV$) is used to constrain
simultaneously the $\alpha_s$ running and the scale (for different
scenarios the limits range from 280--620~$\GeV$).
In~\cite{Ho:2014dha} the CMS
measurements~\cite{Chatrchyan:2013txa,CMS:2014mna}s are used to
constrain a scenario where direct detection of supersymmetric
particles is impossible, but their existence can be manifested by
precise measurements of the strong coupling running at $\TeV$ scales.

\section{Outlook}

In this paper, an analysis of the CMS measurements of the strong
coupling running is combined with a ``traditional'' gauge coupling
unification analysis. This approach places powerful constraints on the
possible scales of new physics and on the parameters around the
unification scale. An MSSM fit without GUT threshold corrections
describes the CMS data well and provides perfect unification with
scales:
$$M_{SUSY}\ =\ 2820\ +670\ -540\ \GeV\ \ \ M_{GUT}\ =\ 1.05 \pm 0.06 \cdot 10^{16}\ \GeV.$$
For zero or small threshold corrections the scale of new physics may
be well within LHC reach.

So far the CMS collaboration has published only the measurements from
the 7~$\TeV$ data. If the LHC experiments benefit fully from the
increased energy and statistics in Run 2 to better control the
systematics and to probe higher energy scales, they can extend
substantially the reach for indirect evidence for the possible scale
of new physics.

\vskip .2cm
\noindent
{\large{\bf{Acknowledgments}}}
\noindent

This work was started in the very productive environment of the
University of Florida High Energy Physics group and completed during a
stay at the LHC Physics Center of the Fermi National Accelerator
Laboratory. The author wants to thank the LPC for the invitation to
visit FNAL and for the kind hospitality.

\end{document}